\documentstyle[12pt,epsf]{article}
\begin{document}
\title{\bf A Model for the Parton Distribution in Nuclei}
\author{J.Ro\.zynek\thanks{e-mail: rozynek@fuw.edu.pl} and
G.Wilk\thanks{e-mail: wilk@fuw.edu.pl}\\[2ex] 
{\it The Andrzej So\l tan Institute for Nuclear Studies}\\
    {\it Ho\.za 69; 00-689 Warsaw, Poland}}
\date{\today}
\maketitle

\begin{abstract}
We have extended recently proposed model of parton distributions in
nucleons to the case of nucleons in nuclei.\\


\end{abstract}

\newpage

Recently a simple model for parton distributions in hadrons has been
presented \cite{EI}. They are derived from a spherically symmetric,
Gaussian distribution, the width of which reflects, via the
Heinserberg uncertainty relation, the hadronic size. Two distinct
parts are distinguished in a hadron: a ``bare'' hadron (identified
with valence quarks and gluons) and hadronic fluctuations (identified
with pions, which are later the source of sea partons). The parton
density distributions were calculated numerically using a 
Monte Carlo technique, and good agreement with deep inelastic
scattering data was reported \cite{EI}.\\ 

In this paper we have used the same method to calculate parton
distributions in nuclei. Preserving the simplicity of the model and 
using the standard nuclear structure we were able to successfully
describe the $F_2(x)$ in nucleus over the whole range of $x$. The
only changes introduced were dictated by the presence of other
hadrons around the one under investigation in a similar fashion as in
our previous work on this subject \cite{RSW}. However, in the present
case we were able to describe also the region of small $x$, which
shows effects of shadowing and therefore is of particular interest to
the physics of heavy ion collisions \cite{HSG}.\\

Let us first recollect the main features of the model \cite{EI},
which we shall follow as close as possible (in particular the choice
of kinematics and notation will be the same). Hadron is visualised
there as a composition of {\it bare hadron} and {\it hadronic
fluctuations}. The former are made out of valence quarks and gluons
and the latter are the source of sea quarks and gluons. They are
formed mainly by the pion. One starts with the hadron at rest in
which frame all partons are supposed to be distributed according to a
spherically symmetric, Gaussian distributions. Such a form is natural
because of a large number of interactions binding partons in the
hadron. The only parameter here is the width of this distribution,
the value of which is expected to be of the order of a few hundred
MeV (both from the Heinserberg uncertainty relation applied to the
hadron size and from the primordial transverse momentum of partons
observed in deep inelastic collisions). This parameter encompasses
also all perturbative QCD effects present due to the initial state
emission and therefore depends on the scale $Q_0^2$ 
\footnote{Actually,
when confronting flavour dependent data the number of parameters is
enlarged because each quark flavour can have its own width; the same
is true for gluons, cf. \cite{EI}. In such situation the percentages
of each species in the momentum sum rule is also a kind of
parameter.}. 
The goal of this approach is not so much the full wave 
function of the hadron, as the probability of finding a single parton
with the four momentum $k$ probed by external current with
four-momentum $q$. Therefore all other partons are treated
collectively as a single remnant with the four momentum $r$. Because
the above prescription provides us only with the three-momentum of
the probed parton it is assumed that the energy component is equal to
the parton (current) mass plus a Gaussian variation with the same
width as above. It means that the parton can be off-shell at the
scale $Q_0^2$ and fluctuates with a life-time given by the nucleon  
radius.\\ 

The reaction takes place in a coordinate system in which the negative
$z$-direction points along the current which probes the hadron. One
uses the light cone momentum fraction x of the parton defined as
$k_{+}=xp_{+}$ (where $p$ is the four-momentum of the hadron). The
final four-momentum of the scattered parton denoted by $j$ must
satisfy the following condition: $0 < j^2 <W^2$ (where $W$ is
invariant mass of the hadronic system). This is equivalent to
$0<x<1$. When masses of quarks are neglected the same condition must
be satisfied by the four-momentum $r$ of the remnant: $ 0< r^2 < W^2$. 
The parton distributions are then calculated by a Monte Carlo code  
\footnote{In our work we have used a simplified version of model
\cite{EI} with the same distributions for all valence quarks 
and without evolution in $Q^2$.  We shall also not
address gluon distributions, which in \cite{EI} were assumed to have
the same basic Gaussian shape as the valence quarks. The presence of
gluons will be accounted for only in the momentum sum rule, where
part of momentum will always be allocated to the gluonic component.
}. 
The momentum $k$ component of the parton to be probed by current with
virtuality $Q^2_0$ and four-momentum $q$ is chosen from the Gaussian
distribution described before. Actually the values of $Q_0^2$ is a
free parameter  expected to be of the order of $1$ GeV$^2$. This
makes it possible to calculate the four-momenta $j$ and $r$. Events
are accepted if the exact kinematical constraints mentioned above are
fulfilled. In this way one obtains as a result the valence parton
distribution  $f_v(x;Q^2_0)$ (``bare'' hadron)
\footnote{One has to remember, however, that it is, the so called,
Bjorken variable $x = x_{Bj} = \frac{Q^2}{2 p\cdot q}$, which is
accessible experimentally whereas \cite{EI} starts from the light
cone target rest frame variable $x = x_{LC} = \frac{k^+}{p^+}$ (where
$p =(M,0,0,0)$) with a fixed resolution $Q^2 = Q_0^2$. Whenever one
is interested in some higher $Q^2$, the QCD evolution in the initial
state must be performed. In this case the $x=x_{Bj}$ picked out of
the proton at $Q^2$ will be smaller than the $x_{LC}=k^+/p^+$ one has
started with at $Q_0^2$. The remaining energy is radiated into the
final state as jets. If the invariant mass squared of all jets in the
final state is $M_X^2$, then $x_{LC}=x_{Bj}\left(1+\frac{M_X^2}{Q^2}
\right)$. In this case, if one only scatters a quark with no further
radiation, then $x_{LC}=x_{Bj} \left(1+\frac{m_q^2}{Q^2}\right)\simeq
x_{Bj}$. This clarifies the limits of applicability of the model used
here.}. 
The sea parton distribution is
given by the pionic component of the nucleon:  $f_s(x;Q^2_0)=\int
\frac{dy}{y} f_{\pi}(y,Q^2_0) f_{pion}(x/y;Q^2_0)$. It is the
convolution of the pion distribution function in the nucleon,
$f_\pi(x;Q^2_0)$, and the parton structure of pion
$f_{pion}(x;Q^2_0)$. This in turn is obtained from the same Gaussian
primodial distribution as used for valence partons. The
characteristic behaviour of the sea partons is then derived from the
pion distribution in the nucleon, which was again parametrized by
Gaussian distribution with a smaller width equal to $0.052$ GeV
reflecting the smallness of the pion mass. The total nucleon
structure function  is then equal to: $F(x;Q^2_0) = f_v(x,Q^2_0)
+f_s(x;Q^2_0)$.\\   

We shall now apply this model to deep inelastic collisions with
nuclei, $l + A \rightarrow l' + X$. As usual in such cases our aim
will be the nuclear structure function $F^A_2(x_A)$, which shows
a characteristic pattern: shadowing at small $x$, followed by
antishadowing at $x\sim 0.1\div0.3$, followed in turn by a very
pronounced deep around $x \simeq 0.7$ and a kind of cumulative effect
for $x\rightarrow 1$. This subject has already long history and a
vast literature (cf. \cite{REF} for review). Our aim is to see if,
and under what conditions, the model proposed in \cite{EI} for free
nucleons can be also applied to nuclei. In the approach presented
here, from many possibilities mentioned in \cite{REF}, we have
selected the picture in which collision proceeds on nucleons bounded
in nuclei, looking for nuclear partonic distributions as a sum of
distributions of bound nucleons.  The corresponding nuclear structure
function can be then written as simple convolution of nuclear and
nucleonic components: 
\begin{equation}
\frac{1}{x_A} F^{A}_{2}(x_{A})\, =\, A \int dy_{A} \int \frac{dx}{x} 
        \delta (x_{A}-y_{A}x) \rho ^{A}(y_{A}) F^{N}_{2}(x) .
        \label{eq:FA}
\end{equation}
Here $F^{N}_{2}(x)$ denotes a nucleon structure function inside the
nucleus and $\rho ^{A}(y_A)$ is the nucleon distribution function in
the nucleus. Variables $x_{A}/A$ and $y_{A}/A$ are the corresponding
Bj\"orken variables for the quarks and nucleons in the nucleus,
respectively (with nucleonic and nuclear longitudinal momenta given
by $p^{+}$ and $P^{+}_{A}$). The presence of nucleus is summarised
here by the nucleon distribution function $\rho$ and the simplest
approach one could think of would be to use available information on
it and keep the nucleonic structure functions unchanged. For example,
in the framework of the relativistic mean field theory (RFM)
\cite{RMF} leptons interact with nucleons with density $\rho^A$,
which are moving in a constant average scalar $U_{S}$ and vector
$U_{V}$ potentials (defined in the rest frame of the nucleus as
$U_{S} = -400 \frac{\rho}{\rho_0}$ MeV and $U_{V} = 300
\frac{\rho}{\rho_0}$ MeV, in both cases $\rho_0 = 0.17$ fm$^{-3}$).
As a result (cf. \cite{B}), in the Relativistic Fermi Gas 
approximation the
following form of the nuclear density function is obtained:  
\begin{equation}
\rho ^{A}(y_A)\, =\, {4\over \rho }\int {d^{4}p\over (2\pi)^{4}} 
S_N(p^o,{\bf p})\left[1\, +\, \frac{p_z}{E^{^{*}}(p)}\right]
          \delta \left(y - \frac{p_o+p_z}{\mu}\right). \label{eq:RHO}
\end{equation}
The $\rho$ is nuclear density (which is different from the density of
infinite nuclear matter $\rho_0$); $S_N=n(p)\delta \left\{ p^0 - [
E^{*}(p) +U_V ]\right\}$ denotes nucleon spectral function with
$n(p)$ being Fermi distribution of nucleon momenta inside the
nucleus, $p\in(0, p_F)$. The corresponding Fermi momentum $p_F
=(3/2\rho\pi^2)^{1/3}$. The nucleon chemical potential $\mu = M_A/A$
(which in RMF can be shown to be $\mu = E^*(p_F) + U_V$) and $E^*(p)=
\sqrt{p^{2}+ m^{*2}}$ with $m^*=m-U_S$. After integration
(\ref{eq:RHO}) simplifies to
\begin{equation}
\rho ^{A}(y_{A})\, =\, {3\over 4}\, \frac{ \left[v^{2}_{A}\, -\,
              (y_{A}-1)^{2}\right]}{v^{3}_{A}} . \label{eq:R}
\end{equation}
Here $v_{A}=p_{F}/E^{^{*}}_{F}$ and $y_A$ takes the values given by
the inequality: $0<(E_{F}^{^{*}} -p_{F})<\mu y_A<(E_{F}^{^{*}} +
p_{F})$. In this way the motion of the nucleon inside the nucleus is
parametrized here by two parameters: nuclear density $\rho$ which
determines Fermi momentum $p_F$ and nucleon chemical potential $\mu$.
Out of these two, the nucleon chemical potential is essentially
constant (and equal to $\mu = 8$ MeV), except for a few very light
nuclei. The nuclear density $\rho$ due to the finite size of the
nucleus can vary from $\rho \simeq 0.1$ fm$^{-3}$ for light nuclei to
$\rho \simeq 0.17$ fm$^{-3}$ for heavy nuclei and nuclear matter.
Taking $\rho \simeq 0.12$ fm$^{-3}$ for the average density of
nucleons in $^{56}$Fe with $p_F=240MeV$ and using (\ref{eq:R}) results in dashed curve
in Fig. 1, which is completely off data. This means that RFM approach
to deep inelastic scattering on nuclei fails, at least in the form
presented here. One could play with nuclear parameters, for example
by increasing chemical potential $\mu$ to make the structure of deep
around $x \sim 0.5 \div 0.7$ more pronounced and more similar to the
experimental data. This would, however, be difficult to justify and
the shadowing/antishadowing structure seen in data at smaller $x$
would remain unexplained. There are then two possibilities: either to
keep nucleon structure functions unchanged and try to change $\rho$
by including correlations \cite{COR} and/or cluster structure 
of nucleons inside the nucleus \cite{CLUS} or to decide 
to change the nucleonic structure functions.
Because the first approach will essentially not affect the shadowing
(or even antishadowing) region of $x$, we have opted for the second
possibility. In what follows we shall therefore use, for nuclear 
structure function, the nuclear density $\rho^A$ as given by eq.
(\ref{eq:RHO}) with $\mu = 8$ MeV and with $U_S=U_V=0$ (i.e.,
$m^*=m$). This is to avoid the possible double counting of nuclear
interactions, which in our model can be expressed either by
potentials $U_{V,S}$ (treated now as free parameters) or by changes
of the free nucleon structure functions discussed below. Because we
have decided (as said above) for the latter possibility we have
consistently set values of potentials equal to zero. It means 
that we are considering here a model of Fermi gas with modified
nucleons.\\ 

Our Monte Carlo calculation for collision on nuclei has been
performed in the same way as in \cite{EI}. The only difference was 
that now the nucleon three-momentum is not fixed but chosen from
distribution $\rho^A(y_A)$ as given by eq. (\ref{eq:RHO}), i.e., we
account for the Fermi motion of nucleons in nucleus. This introduces
a kinematical medium effect produced by Lorentz transformation 
of the parton momenta from the nucleonic to the nuclear rest frame.
This Fermi motion of nucleons also affects invariant mass $W$, four
momentum $r=p-j$ and Bjorken variable $x=k_{+}/p_{+}$ and in this way
influences the calculated structure function $F_{2}^{B}(x,p)$ of the
bound nucleon making it momentum dependent. Analytically it would
mean that one extends our simple convolution formula (\ref{eq:FA})
replacing $F^N_2(x)$ by momentum dependent $F_2^B(x,p)$ and
integrates over momentum $p$, i.e., instead of (\ref{eq:FA}) one uses
formula
\begin{equation}
 F^{A}_{2}(x_{A})\, =\, {4A\over \rho } \int dy_{A} 
                     \int {d^{4}p\over (2\pi)^{4}} \, S_N(p)\,
                     \left[1\, +\, \frac{p_z}{E^{^{*}}(p)}\right]
          \delta \left(y_A-{p_+\over \mu}\right)\, 
          F^{B}_{2}\left({x_A\over y_A},p\right) .  \label{eq:LAST}
\end{equation}
The other modifications mentioned before must be made by some
suitable changes in parameters of the original nucleonic structure
functions. Because we are not differentiating here between partons of
different flavour there are only three such parameters, which can be
argued to be affected by nuclear medium: the widths of the original
gaussians (i.e., transverse primordial distributions), the same and
equal $\sigma_q=0.18$ GeV for both valence quarks (``bare'' hadron)
and for nucleonic pions (sea quarks) and the width of the pionic
distribution in nucleon equal $\sigma_{\pi}=0.052$ GeV. It turns out
that, in order to get reasonable description of data, it is enough to
modify only $\sigma_q$ (keeping it again the same for both types of
quarks) by decreasing it to the value $\sigma_q = 0.165$ GeV and
centering pionic energy distribution not at the pion mass $m_{\pi} = 0.14$
GeV but at $m_{\pi} = 0$. This can be seen in Fig. 1 where we present
results of calculations of ratio $R(x) =f_2^A(x)/f_2^D(x)$ for
$^{56}$Fe (performed using average value of nuclear density $\rho
=0.12fm^{-3}$, $p_F=240MeV$). The mentioned results are presented by open squares.
Such a change in the width of initial partonic distribution can be
naturally explained by the expected swelling of nucleons in nuclear
matter \cite{RSW,REF}. Smaller spread of momenta caused by the
internucleonic interactions corresponds, due to the uncertainty
relation, to bigger spread in coordinate space (the ``vacuum'' in
nucleus is obviously different from that outside the nucleus).
Although the idea is similar to what was proposed a long time ago in
\cite{RSW} the effect of swelling is this time modelled not by
changing the bound nucleon mass but by changing the intrinsic motion
of partons inside such nucleon. What concerns the change made in the
center of the energy distribution of nucleonic pions one can argue,
following for example \cite{R}, that in nuclear matter such pions,
interacting strongly with nuclear matter and coupled to the nuclear
collective modes, seem to have indeed vanishing effective mass.  
In any case such change is essential in getting agreement with
data. With only swelled nucleon and without decreasing $m_{\pi}$
(understood as a position of maximum in the energy distribution of
nucleonic pions) results in curve denoted by open diamonds in Fig. 1.
The other parameters were left unchanged (in particular, the fraction
of momentum carried by the sea quarks is left $7.7\%$ as in
\cite{EI}). Contrary to what has been presented in \cite{RSW} this
time we have also reproduced (at least partially) the
shadowing/antishadowing region of $x$. It has to be admited here,
however, that there is a price in obtaining these results (see open
squares in Fig. 1). The momentum sum rule is now violated by $\sim
2\%$, which in our case means that this fraction of nucleonic momenta
is taken by gluons (in addition to what they already had in free
nucleons). If we strictly impose the momentum sum rule (i.e., if we
assume that there is no additional transfer of the momenta to
gluons), it results in changing the momentum carried by quarks to
$9.7\%$ and spoils the agreement with data in the vicinity of the
antishadowing region (cf. Fig. 1, open triangles). \\     

To conclude, we have demonstrated that the simple model of partonic
distributions in nucleons developed recently in Ref. \cite{EI} can
also sucessfully describe partonic distributions in nuclei by:
$(i)$ using standard description of nucleonic Fermi motion, $(ii)$
changing by small amount (about $10\%$) the Gaussian widths of the
initial partonic components and $(iii)$ centering the energy
distribution of nucleonic pions on $m_{\pi} = 0$ instead $0.14$ GeV.
These changes correspond to the valence quark distributions in nuclei
being shifted towards the lower values of x and the sea quark
distribution being relatively spread out and shifted towards the
higher values of x. Also nucleons in nucleus allocate a bit more
momentum in the gluonic component. Altogether, by these simple means
a good agreement with data was obtained for the whole range of $x$
except of really small values of $x \leq 0.06$ where such small
changes are definitely not sufficient and convolution model is hardly
justifiable \cite{REF}. We consider it  remarkable that by changing
only one parameter describing the primordial diffusiveness of partons
(in momentum space) in the free nucleon and shifting pionic
distribution in way suggested by nuclear matter calculations we were
able to describe experimental data in a fairly reasonable way in a
very broad region of $x$. Actually, our results are quite robust to
changes in initial values of parameters describing partonic
distributions in free nucleon (for example, a $20\%$ change in the
quark distribution in pion does not change results presented in Fig.
1). Our analysis is obviously simplified, as we have not
differentiated between quark flavours or accounted for the presence
of gluons only via the momentum sum rule. Aso our convolution
approach, cf. eq. (\ref{eq:FA}) is not suitable for small values of
$x$. But already at this level it shows that model \cite{EI} is
highly suitable for application to nuclear partonic distributions
because of simplicity and cogency of the parametrization chosen
there.\\  

\vspace{4mm}

\vspace{4mm}
\noindent
Acknowledgements:\\
We are grateful to Prof. L.\L ukaszuk for many helpful discussions.
This work has been supported in part by the Committee for Scientific
Research, grant 2-P03B-048-12.\\

\newpage
\noindent
{\bf Figure Captions:}\\

\begin{itemize}
\item[{\bf Fig. 1}] Results for $R(x)=F^{A}_{2}(x)/F^{D}_{2}(x)$ for 
                    $^{56}Fe$. Data are from \cite{DATA}
                    - full diamonds and from \cite{DATA1,DATA2} 
                    - full circles. See text for explanations.
\end{itemize}

\newpage
\begin{figure}
\epsfxsize=18cm
\epsfbox{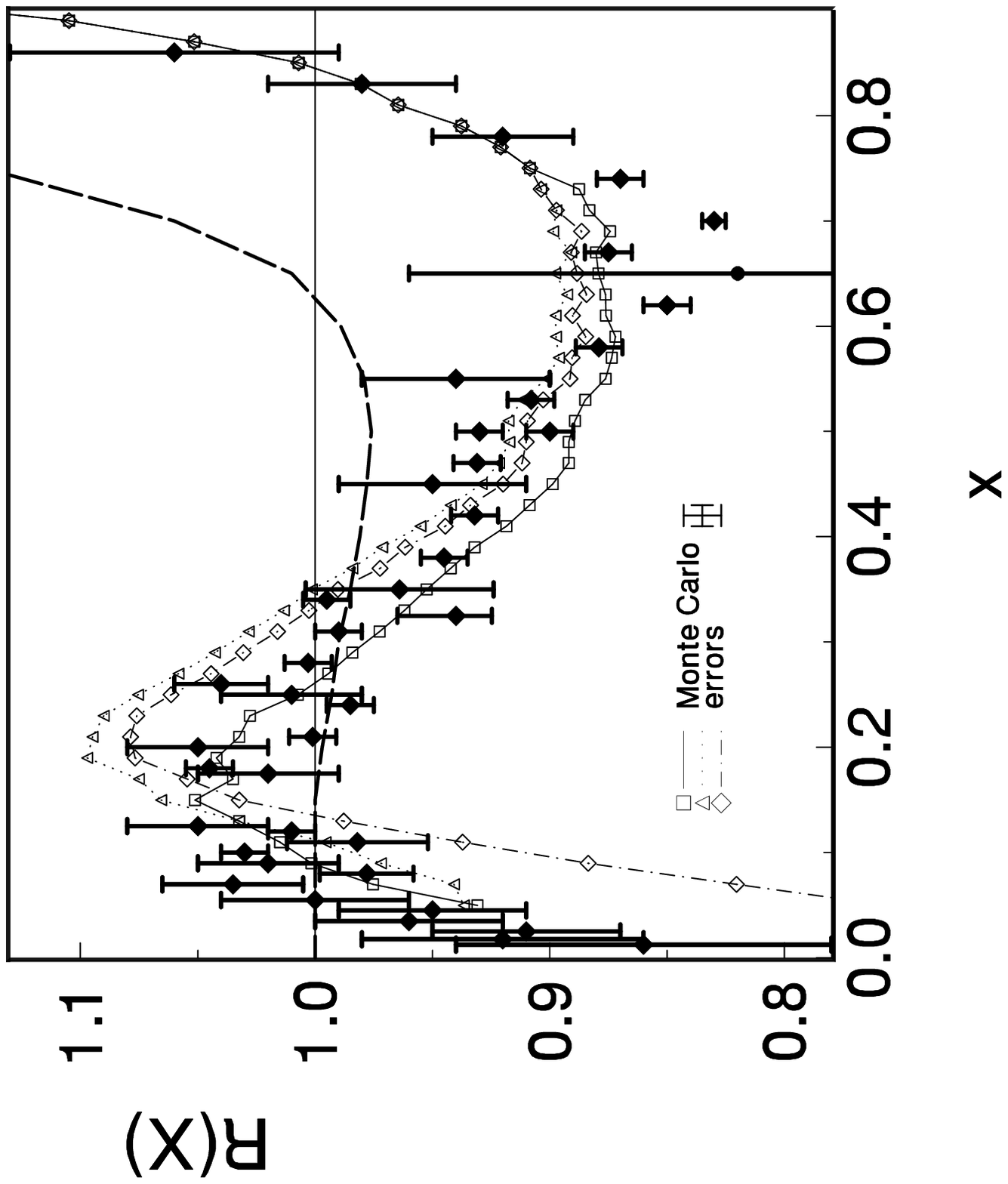}
\end{figure}

\end{document}